\documentclass[conference,10pt]{IEEEtran}
\IEEEoverridecommandlockouts
\usepackage{cite}
\usepackage{amsmath,amssymb,amsfonts}
\usepackage{bm} 
\usepackage{graphicx}
\usepackage{textcomp}
\usepackage{xcolor}
\usepackage{multirow}
\usepackage[ruled]{algorithm}
\usepackage{algpseudocode}
\usepackage{subcaption,lipsum}
\usepackage{booktabs,tabularx}
\usepackage{tikz}
\usepackage[utf8]{inputenc}
\usepackage{pgfplots} 
\usepackage{pgfgantt}
\usepackage{pdflscape}
\pgfplotsset{compat=newest} 
\pgfplotsset{plot coordinates/math parser=false}
\usepackage{utfsym}

\def\BibTeX{{\rm B\kern-.05em{\sc i\kern-.025em b}\kern-.08em
    T\kern-.1667em\lower.7ex\hbox{E}\kern-.125emX}}

\begin{document}

\title{Simultaneous Communication and Tracking using Fused Bistatic Measurements
}

\author{\IEEEauthorblockN{Avinash M and Srikrishna Bhashyam$^{1}$\thanks{Avinash M was a Masters' student at the Dept. of Electrical Engineering, IIT Madras. This work was done at IIT Madras as part of his Masters' project. He is currently with CABS, DRDO, Bangalore, India. }}
\IEEEauthorblockA{$^1$Dept. of Electrical Engineering \\
IIT Madras, Chennai 600036, India\\
ee22m002@smail.iitm.ac.in, skrishna@ee.iitm.ac.in}
}

\maketitle

\begin{abstract}
In this paper, we propose a {\em bistatic sensing}-assisted beam tracking method for simultaneous communication and tracking of user vehicles navigating arbitrary-shaped road trajectories. Prior work on simultaneous communication and tracking assumes a colocated radar receiver at the transmitter for sensing measurements using the reflected Integrated Sensing and Communication (ISAC) signals in the mmWave band. Full isolation between transmitter and receiver is required here to avoid self-interference. We consider the bistatic setting where the sensing receivers are not colocated and can be realized in practice using traditional half-duplex transmit or receive nodes. First, we process the echoes reflected from the vehicle at multiple multi-antenna nodes at various locations, facilitating estimation of the vehicle's current position. Then, we propose selection criteria for the estimates and a maximum likelihood (ML) fusion scheme to fuse these selected estimates based on the estimated error covariance matrices of these measurements. This fusion scheme is important in bistatic and multistatic settings as the localization error depends significantly on the geometry of the transmitter, target, and receiver locations. Finally, we predict the vehicle's next location using a simple kinematic equation-based model. Through extensive simulation, we study the average spectral efficiency of communication with a moving user using the proposed simultaneous communication and tracking scheme. The proposed fusion-based scheme achieves almost the same average spectral efficiency as an ideal scheme that knows the exact trajectory. We also show that the proposed scheme can be easily extended to systems with Hybrid Digital-Analog architectures and performs similarly even in these systems.
\end{abstract}


\section{Introduction}
\label{sec:intro}
In Vehicle-to-Everything (V2X) – 6G use case scenarios, effective communication between the Road Side Unit (RSU) and mobile vehicles demands high data rates. To address limited bandwidth availability and achieve optimal spectral efficiency, Multiple Input Multiple Output (MIMO) transceivers operating in mm-wave frequencies are envisioned \cite{Beam_Manage_1}. Efficiently managing narrow beams toward mobile users in a high-mobility environment in mm-wave systems necessitates effective beam tracking \cite{Beam_train_2}. Traditional techniques, such as beam sequence training, incur significant overhead and are ill-suited for mobile scenarios \cite{Beam_train_2}. 
Simultaneous communication and beam tracking using a monostatic colocated transmitter-receiver configuration has been recently proposed in \cite{Beam_space_processing}. In \cite{Beam_space_processing}, the echoes of the communication/ISAC waveform from the user/target are processed to track the user and point the beam appropriately.  However, achieving isolation between the transmitter and receiver in this monostatic setting is difficult and may result in residual self-interference \cite{SI_prob_sol}. 

In this paper, we extend the monostatic simultaneous communication and tracking setting in \cite{Beam_space_processing} to bistatic sensing-assisted communication and tracking. We also allow for the possibility of multiple sensing receivers as outlined in \cite{SI_prob_sol}. The bistatic setting with a separate transmitter and receiver addresses the self-interference problem and has potential advantages like better Radar Cross Section (RCS) detection, increased path diversity gain, and increased sensing range by mobile receiver \cite{GDOP}.
However, the positional error of a bistatic system is sensitive to  the geometry of the transmitter (TX), Receiver (RX), and Target  \cite{GDOP}. This is addressed in our work using appropriate error covariance estimation, and fusion of estimates from multiple receivers based on selection of good receivers.
We propose a bistatic sensing-assisted beam-tracking method in this setting for simultaneous communication and tracking. This method consists of: (1) a location estimation step at each receiver, (2) error covariance estimation and selection criteria for selecting the estimates from different receivers (3) a maximum likelihood fusion scheme to combine the selected location estimates based on estimated covariance matrices, and (4) a prediction step that predicts the location at the next time step using a kinematic model.
Previous approaches to fuse multiple position estimates or utilize multiple measurements \cite{GDOP},\cite{Why_mono_better}, have typically assumed fixed variances for measurement noise at each receiver. In contrast, we determine the error covariance matrix for each location estimate using Cramer-Rao Lower Bound (CRLB) expressions \cite{MUSIC}\cite{C_tau_book} that incorporate the dependence of the error on the TX-Target-RX geometry and Signal to Noise Ratio (SNR). As in \cite{Beam_space_processing}, due to good suitability as an ISAC waveform, we use the Orthogonal Frequency Division Multiplexing (OFDM) waveform.
Using simulations on an arbitrary road trajectory, we show that: 
(1) the proposed simultaneous communication and beam tracking scheme using bistatic estimates and maximum likelihood fusion can approach the spectral efficiency of a communication scheme with perfect prior location information, (2) the proposed scheme can also be applied to systems with hybrid digital analog architectures that allow for lower complexity implementation.

\section{System Model}
\label{sec:model}
\begin{figure}[t]
\centering
\input{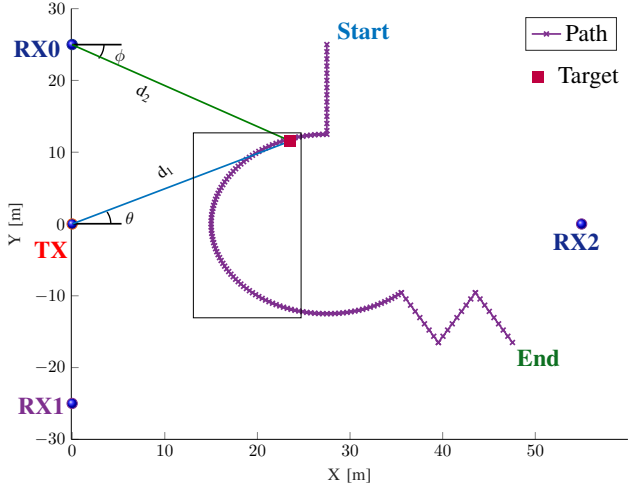}
\caption{True path of the Target}
\label{fig:path2}
\end{figure}

Consider a Transmitter (TX) with a Uniform Linear Array (ULA) consisting of $N_{t}$ isotropic antennas communicating to $K$ communication users while simultaneously beamforming and tracking them as in \cite{Beam_space_processing}.  At every measurement epoch, we assume that the coarse Angle of Departure (AoD) of each user is known to the TX using a beam alignment method (for a new user) or using a predicted location from previous estimates for the existing users. We consider OFDM frames to be the ISAC waveform. The reflected echoes from the Users are received by multiple receivers (RX) each with $N_{r}$ antenna ULA system and not colocated with the TX. The model in \cite{Beam_space_processing} considers a monostatic scenario (with co-located TX and RX). However, we have extended this to a bistatic configuration. It is assumed that the TX and receivers are synchronized and connected to a central Processor (CP). The receivers know the transmitted waveform to perform matched filtering. See Fig. \ref{fig:path2} for a sample scenario with 3 receivers RX0, RX1 and RX2.

\subsection{Received Signal Model}
We have adopted the same mm-Wave Radar model for the echoes and used OFDM as the ISAC waveform, as detailed in \cite{Beam_space_processing}. The main difference is that the AoA and AoD for each user are different since the TX and RX are not colocated. The received echo at each RX in the absence of noise is 
\begin{equation}
\begin{split}
        \boldsymbol{y}(t) &= \sum_{k = 1}^{K} h_{k} ~ \boldsymbol{b}(\phi_{k}) ~ \boldsymbol{a}^{H}(\theta_{k}) ~\boldsymbol{s}(t - \tau_{k}) ~e^{j2\pi\gamma_{k} t}, \\ 
\end{split}
\end{equation}
where $\boldsymbol{s}(t)$ is the transmitted OFDM frame with $N$ symbols and $M$ subcarriers as follows 
\begin{equation}
\label{tx_signal}
     \boldsymbol{s}(t) = \sum_{k = 1}^{K} 
 \boldsymbol{f}(\Hat{\theta}_{k}) \sum_{n = 0}^{N-1} \sum_{m = 0}^{M-1} \zeta_{k}[n,m] p_{n,m}(t),
 \end{equation}
 and the OFDM pulse is
 \begin{equation}
     p_{n,m}(t) = \text{rect}\left(\frac{t-nT_{o}}{T_{o}}\right)~e^{j2\pi m \Delta f (t-T_{cp}-nT_{o})}.
 \end{equation}
 Here $\text{rect}(x)$ is $1$ when $0~ \leq x \leq 1$ and $0$ elsewhere, $\Delta f$ is the sub-carrier frequency spacing, $T_{cp}$ is the cyclic prefix, $T_{o} = (1/\Delta f) + T_{cp} $ and $\zeta_{k}[n,m]$ is the transmitted QPSK symbol for the $k^{th}$ user during the $n^{th}$ symbol on the $m^{th}$ subcarrier with power $ P_{T}/K$. $\{\Hat{\theta}_{k}\}$ is the set of predicted AoDs, $\boldsymbol{f}(\hat{\theta}_k) = \frac{1}{\sqrt{N_{t}}}~\boldsymbol{a}(\hat{\theta}_{k})$ is the unit norm beam pointing towards the $k^{th}$ user. We consider ULAs with $\lambda/2$ spacing, i.e., $ [\boldsymbol{a}(\theta_{k})]_i = e^{j{\pi(i-1)\sin{\theta_{k}}}},
[\boldsymbol{b}(\phi_{k})]_i = e^{j{\pi(i-1)\sin{\phi_{k}}}}$. $\tau_{k},\gamma_{k},\phi_{k},\theta_{k}$ are respectively the total delay (TX-User-RX), doppler, Angle of Arrival (AoA) and Angle of Departure (AoD) of the $k^{th}$ user. $h_{k}$ is the complex channel reflection coefficient  with squared magnitude 
$
   \left| h_{k} \right|^2 = \frac{\lambda^2 \sigma_{\text{rcs}}}{(4\pi)^3 d_{k1}^2 d_{k2}^2}
$ and Uniform random phase $\in [-\pi,\pi] $, where $d_{k1},d_{k2}$ is the TX-User and RX-User distance and $\sigma_{rcs}$ is the Radar cross section (RCS) respectively. The wavelength is $\lambda = c/f_{c}$, $f_{c}$ is the carrier frequency. We assume that $T_{cp} \geq d_{max}/c$ and $\Delta f \gg \gamma_{max}$, where $d_{max}$ and $\gamma_{max}$ is the maximum delay and doppler which are determined by maximum sum range ($d_{k1}+d_{k2}$) and maximum user velocity $v_{max}$.

Under the assumption that the users are well separated and with good estimation of AoD, in massive MIMO, we can assume $ \left| \boldsymbol{a}^{H}(\theta_{l}) \boldsymbol{f}(\hat\theta_{k}) \right| \approx 0 $ for $l \neq k$ \cite{Beam_space_processing}, resulting in 
\begin{multline}
        \boldsymbol{y}(t) 
    \approx \sum_{k = 1}^{K} h_{k} \boldsymbol{b}(\phi_{k}) \boldsymbol{a}^{H}(\theta_{k}) \boldsymbol{f}(\Hat{\theta}_{k})\\ \sum_{n = 0}^{N-1} \sum_{m = 0}^{M-1} \zeta_{k}[n,m] p_{n,m}(t-\tau_{k}) e^{j2\pi\gamma_{k} t}.
\end{multline}
After standard OFDM processing \cite{OFDM_processing} and including noise, the sampled $N_r \times 1$ signal at each RX is given by
\begin{multline}
    \bm{y}[n,m] = \sum_{k = 1}^{K} h_{k}\boldsymbol{b}(\phi_{k}) \boldsymbol{a}^{H}(\theta_{k}) \boldsymbol{f}(\hat{\theta}_{k})\\
    \zeta_{k}[n,m] ~e^{j2\pi (nT_{o}\gamma_{k} - m\Delta f \tau_{k})}
    + \bm{z}[n,m],
\end{multline}
where $\bm{z}[n,m] \in \mathbb{C}^{N_{r}}$ is circular white Gaussian noise with variance $\sigma^{2} = N_{o} BW$, $N_{o}$ is the noise power spectral density and $BW = M\Delta f$ is the bandwidth.
\section{Proposed Beam Tracking algorithm}
In this section, we first present how the position estimates are obtained at each receiver. Then, (1) we derive estimates of the error covariance matrix for the position estimates, (2) we propose a selection criteria for choosing a subset of estimates and (3) we develop a maximum likelihood fusion technique based on the estimated error covariance matrices for the fusion of selected estimates. 
\subsection{Position estimation}
\label{subsection:Para_est}
Position is estimated at each receiver as follows.
\subsubsection{Angle of Arrival estimation AoA}
\label{sub:est_AoA}
In each measurement epoch, we obtain the sample covariance matrix from the $NM$ samples of $\bm{y}[n,m] \in \mathbb{C}^{N_{r}} $.
Then, the AoA estimate of user $k$, $\tilde{\phi}_{k}$, is estimated at each receiver using MUltiple SIgnal Classification (MUSIC) \cite{MUSIC}. 
\subsubsection{Estimation of Delay and Doppler frequency}
\label{sub:est_d_doppler}
Based on the estimated AoA $\Tilde{\phi}_{k}$ for the user $k$, we can construct the receive beamformer $\boldsymbol{w}(\tilde{\phi}_{k}) = \frac{1}{\sqrt{N_{r}}}~\boldsymbol{b}(\Tilde{\phi}_{k})$ and get
 $   \Tilde{y}_{k}[n,m] = \boldsymbol{w}(\tilde{\phi}_{k})^{H} ~ \boldsymbol{y}[n,m].$
Since the RX knows the transmitted data, we can get
\begin{equation}
    \begin{split}
    \Tilde{r}_{k}[n,m] &= \frac{\zeta_{k}[n,m]}{\boldsymbol{w}(\tilde{\phi}_{k})^{H} ~ \boldsymbol{w}(\tilde{\phi}_{k}) } ~ \Tilde{y}_{k}[n,m].
    \end{split}
\end{equation}
Now, from the $N \times M$ matrix with elements $\Tilde{r}_{k}[n,m]$, we can estimate the delay and doppler ($\Tilde{\tau}_{k}$ and  $\Tilde{\gamma}_{k}$ respectively),  by finding the peak in the  2D Delay-Doppler grid \cite{OFDM_radar_est} choosing an appropriate oversampling factor $S$ for the FFT. 
\subsubsection{Estimation of Position}
\label{sub:position_est}
The bistatic sum range $\Delta R \triangleq d_{1} + d_{2}$ is obtained from the estimated delay $\Tilde{\tau}_{k}$ as $\Delta R = c \Tilde{\tau}_{k}$. The distance $d_2$ for user $k$ is found using the bistatic equation \cite{GDOP} as follows 
\begin{equation}
\label{equ:d2}
    d_{2} = 0.5 \frac{(\Delta R)^{2} - L^{2}}{\Delta R -  L\sin{\Tilde{\phi}_{k}}},
\end{equation}
where $L$ is the distance between the TX and RX. Finally, the estimated position of the target $k$, ($\tilde{x}$, $\tilde{y}$), is given by $(x_r + d_2\cos{\Tilde{\phi}_{k}},y_r +d_2\sin{\Tilde{\phi}_{k}})$, where $(x_r, y_r)$ is the location of the receiver. Unlike the monostatic configuration, the sensitivity of localization error in a bistatic setting depends on the TX-target-RX geometry due to the non-linearity in (\ref{equ:d2}).

\subsection{Measurement Error Covariance and its Estimate}
\label{sec:covest}
Let the true position of the target, say user $k$, be $\boldsymbol{\alpha} = [x~y]^{\top}$ and the position estimate at receiver RX$_i$ be $\boldsymbol{T}_{i} = [\tilde{x}_{i} ~\tilde{y}_{i}]^{\top}$.
The position estimate $\boldsymbol{T}_i$ at receiver $i$ is obtained as described in Sec. \ref{sub:position_est} by transforming $\boldsymbol{Z}_i = [\tilde{\tau}^i ~ \tilde{\phi}^i]^{\top}$, where $\tilde{\tau}^i$ is the delay estimate and $\tilde{\phi}^i$ is the AoA estimate at the $i^{th}$ receiver.

The error covariance matrix for $\boldsymbol{Z}_i$ is determined as $C_{Z_i} = \begin{bmatrix}
        C_{\tau} & 0 \\
       0 & C_{\phi} 
    \end{bmatrix}$, where $C_{\phi}$ is the Cramer-Rao lower bound (CRLB) for the measurement error variance in AoA given by (using \cite[Thm. 4.1]{MUSIC})  
\begin{align}
    C_{\phi} &= \frac{1}{~ \rho_{o} NM}(1+\frac{1}{N_{r} \rho_{o}}) \times \frac{6}{N_{r}(N_{r}^{2}-1)},
\end{align}
and $\rho_{0}$ is the per-antenna received Signal to Noise ratio (SNR) given by \cite{Geo_aware_ISAC}
\begin{align}
\label{rho_0_actual}
    \rho_{0} &= \frac{P_T \left |h_{k}\right|^2 \left| \boldsymbol{a}^{H}(\theta_{k}) \boldsymbol{f}(\hat{\theta}_{k}) \right|^2 }{K \sigma^{2}}.
\end{align}
$C_{\tau}$ is the CRLB for the measurement error variance in $\tau$, which is given by  \cite{C_tau_book}
\begin{align}
     C_{\tau} &= \frac{3}{2\pi^{2} S^{2} (M\Delta f)^{2}  \rho_{1} N T_{0}},
\end{align}
where $\rho_{1} = \left| \boldsymbol{w}^{H}(\tilde{\phi}_k)\boldsymbol{b}(\phi_k)\right|^{2} \rho_{0}$ , is the beamformed SNR \cite{Geo_aware_ISAC}, and $S$ is the FFT oversampling factor in Section \ref{sub:est_d_doppler}.

Since the true $\theta_{k}$ and $\phi_k$ are unknown, $\rho_0$ and $\rho_1$ are unknown, and have to be estimated. We use the eigenvalues of the covariance matrix $\bm{R} = \frac{1}{NM}\sum_{n = 0}^{N-1} \sum_{m = 0}^{M-1} \bm{y}[n,m] \bm{y}^{H}[n,m]$ obtained in MUSIC to estimate $\rho_0$ as follows. Let the decending ordered eigenvalues of $\bm{R}$ be $\lambda_1, \lambda_2, \hdots, \lambda_{N_r}$. $\lambda_1$ is used as an estimate of the received signal power across all the antennas. The noise variance $\sigma^2$ is estimated as 
\[
\hat{\sigma^2} = \frac{1}{N_r -1} \sum_{j = 2}^{N_r} {\lambda_j}.
\]
Therefore, we have the estimate of $\rho_0$ to be:
\begin{align}
\label{rho_0_est}
    \rho^{est}_{o} &= \frac{\lambda_1}{N_r \hat{\sigma^2}}.
\end{align}
$\rho_{1}$ is estimated using the beamformed received samples as:
\begin{align}
\label{roh_1_est}
   \rho^{est}_{1} &= \frac{P_T }{K \hat{\sigma^{2}}} \left| \frac{1}{NM} \sum_{n = 0}^{N-1} \sum_{m = 0}^{M-1} \left| 
  \frac{\Tilde{y}_{k}[n,m]}{\zeta_{k}[n,m]}  \right| \right|^{2}.
\end{align}
 The estimated error covariance matrix $\hat{C}_{Z_i} = \begin{bmatrix}
        \hat{C}_{\tau} & 0 \\
       0 & \hat{C}_{\phi} 
    \end{bmatrix}$, where $\hat{C}_{\tau}$ and $\hat{C}_{\phi}$ are obtained by using the estimates $\rho^{est}_{0}$ and $\rho^{est}_{1}$, respectively, in place of $\rho_0$ and $\rho_1$.
Now, as in \cite{GDOP}, we use the linear approximation for the transformation from $\boldsymbol{Z}_i$ to $\boldsymbol{T}_i$ assuming a small measurement error. Let $d\boldsymbol{Z} = [d\tau ~ d\phi]^{\top}$ and $d\boldsymbol{Z} = \boldsymbol{J} \Delta\boldsymbol{T}$, where $\Delta\boldsymbol{T} = [dx ~dy]^{\top}$ and $\boldsymbol{J}$ is the Jacobian matrix.
Then, the error covariance matrix  $\boldsymbol{\Sigma}_i$ for the position estimate at receiver $i$ is given by  
 $   \boldsymbol{\Sigma}_i  =\boldsymbol{B}  \hat{C}_{Z_i} \boldsymbol{B}^{\top}$,
where $\boldsymbol{B} = (\boldsymbol{J}^{\top} \boldsymbol{J})^{-1} \boldsymbol{J}^{\top}$.

\subsection{Selection Criteria}
From the derivation of the error covariance matrix estimate above, we observe that the positional error depends on (1) measurement error in delay and AoA estimates, and (2) the scenario geometry captured by the Jacobian matrix $\boldsymbol{J}$.
The measurement error depends upon the received SNR, and the sensitivity dependence on scenario geometry is due to the nonlinearity of the bistatic equation (\ref{equ:d2}). In the monostatic setting, the error depends mainly only on the received SNR. Therefore, unlike the monostatic setting, in the bistatic setting, it is important to estimate the error covariance at each position of the target and then select the receivers to be used for fusion. 

We use $\text{GDOP}_i = \sqrt{\text{trace}(\boldsymbol{\Sigma}_i)}$, called the Geometric Dilution of Precision (GDOP) \cite{GDOP}, to select receivers. During each measurement interval, we select $N_{\text{sel}}$ receivers out of the available $N_{\text{RX}}$ receivers. We choose the receivers with lower GDOP. In the simulation section, we select 2 receivers out of 3 available receivers. 

Next, we propose a fusion method to fuse the estimates from the selected receivers based on the estimated error covariance matrices and the maximum likelihood principle.

\subsection{Maximum Likelihood-based Fusion}
\label{sec:fusion}
For simplicity, we first describe the details of the fusion method for the two receiver cases. The general formula for combining measurements from $N_{\text{RX}}$ receivers is obtained similarly and is given at the end. As defined earlier, let the true position of the target be $\boldsymbol{\alpha} = [x~y]^{\top}$ and the position estimates of the two receivers RX0 and RX1 be $\boldsymbol{T}_{0} = [x_{0}^{t} ~y_{0}^{t}]^{\top}$ and $\boldsymbol{T}_{1}=[x_{1}^{t} ~y_{1}^{t}]^{\top}$ respectively. We model $\boldsymbol{T}_{0} = \boldsymbol{\alpha} + \boldsymbol{N}_{0}$, $\boldsymbol{T}_{1} = \boldsymbol{\alpha} + \boldsymbol{N}_{1}$, where $\boldsymbol{N}_{0}$ and $\boldsymbol{N}_{1}$ are independent Gaussian noise vectors with covariance matrices $\boldsymbol{\Sigma}_{0}$ and $\boldsymbol{\Sigma}_{1}$, respectively. 
We have already estimated $\boldsymbol{\Sigma}_{0}$ and $\boldsymbol{\Sigma}_{1}$ as described in Section \ref{sec:covest}. Then, the likelihood of $\boldsymbol{T}_{0}$ and $\boldsymbol{T}_{1}$ for a given $\boldsymbol{\alpha}$ is
\begin{equation}
\begin{split}
    f(\boldsymbol{T}_{0},\boldsymbol{T}_{1}/\boldsymbol{\alpha})) &= 
    \frac{ e^{\frac{-1}{2}((\boldsymbol{T}_{0}-\boldsymbol{\alpha})^{T}\boldsymbol{\Sigma}_{0}^{-1}(\boldsymbol{T}_{0}-\boldsymbol{\alpha})  + (\boldsymbol{T}_{1}-\boldsymbol{\alpha})^{T}\boldsymbol{\Sigma}_{1}^{-1}(\boldsymbol{T}_{1}-\boldsymbol{\alpha}))}}{(2\pi)^{2}\sqrt{\det{\boldsymbol{\Sigma}_{0}\det{\boldsymbol{\Sigma}_{1}}}}}.
\end{split}
\end{equation}
Solving   
$\underset{\boldsymbol{\alpha}}{\operatorname{argmax}} ~\log{f(\boldsymbol{T}_{0},\boldsymbol{T}_{1}/\boldsymbol{\alpha})},$
the fused estimate of the target’s position is given by 
\begin{equation}
\label{eq:fusion_est}
    \Tilde{\boldsymbol{\alpha}} = ( \boldsymbol{\Sigma}_{0}^{-1}+\boldsymbol{\Sigma}_{1}^{-1})^{-1}(\boldsymbol{\Sigma}_{0}^{-1}\boldsymbol{T}_{0} + \boldsymbol{\Sigma}_{1}^{-1}\boldsymbol{T}_{1}).
\end{equation} 
Extending the above results for $N_{\text{RX}}$ independent receivers, we get 
the fused estimate is to be
\begin{equation}
\label{equ:fuse_meas}
    \Tilde{\boldsymbol{\alpha}} = \left(\sum_{i = 0}^{i = N_{\text{RX}}-1} \boldsymbol{\Sigma}^{-1}_{i}\right)^{-1} \left( \sum_{i=0}^{i=N_{\text{RX}}-1} \boldsymbol{\Sigma}^{-1}_{i} \boldsymbol{T}_{i}   \right),
\end{equation} 
and the positional error covariance matrix of the fused estimate is given by $\boldsymbol{\Sigma}_{fused} = (\sum_{i = 0}^{i = N_{\text{RX}} - 1} \boldsymbol{\Sigma}^{-1}_{i})^{-1}$. 
\begin{algorithm}[!hb]
\caption{Beam Tracking Algorithm for user $k$ at $[x~~y]^T$}
\label{Algo:1_bi_ISAC}
\begin{algorithmic}[1] 
\Require Refresh period $\Delta T$, $N_{\text{RX}}$, $N_{\text{sel}}$, Initialize predicted AoD $\{\hat{\theta}^0\}$, $l \gets 0$,  and miss-count$_{k}$   $\gets$ 0.
\State \textbf{loop}
\State Transmit beamformed OFDM frames in the predicted direction as in (\ref{tx_signal}) based on predicted AoD $\{\hat{\theta}^{l}\}$.
\State $N_{\text{invalid}}$ $\gets$ 0
\For {each receiver $i$} 
\State Estimate position $\boldsymbol{T}_i^l = [\Tilde{x}_{i}^l 
~\Tilde{y}_{i}^l]^{\top}$.
\If{$\| \boldsymbol{T}_i^l - \boldsymbol{T}^{l}_{\text{pred}} \|> \beta $ } 
\State  $N_{\text{invalid}}$ $\gets$ $N_{\text{invalid}}$ + 1
\Else 
\State miss-count$_{k}$  $\gets$ 0 
\State Estimate GDOP of the receiver. 
\EndIf
\EndFor

\If{$N_{\text{invalid}}$ $=$ $N_{\text{RX}}$}
\State miss-count$_{k}$  $\gets$ miss-count$_{k}$ +1
\State $\boldsymbol{T}^l = \boldsymbol{T}^{l}_{\text{pred}}$ 
\Else
\State Select $\min(N_{\text{sel}},N_{\text{RX}}-N_{\text{invalid}})$ valid receivers with the lowest GDOP
\State Fuse estimates from all the selected receivers using Sec. \ref{sec:fusion} to get $\boldsymbol{T}^l = [\tilde{x}^l~~\tilde{y}^l]^T$.
\EndIf
 \State Predict position $\boldsymbol{T}_{\text{pred}}^{l+1} = [\hat{x}_{l+1},\hat{y}_{l+1}]$, $\hat{\theta}_{l+1}$ at time $l+1$ using the estimates $\boldsymbol{T}^l$, $\boldsymbol{T}^{l-1}$ and $\boldsymbol{T}^{l-2}$.

\State $l\gets l+1$
\State \textbf{End loop}
\end{algorithmic}
\end{algorithm}

\subsection{Prediction model and Beam Tracking}
The proposed Beam Tracking algorithm is presented in Algorithm \ref{Algo:1_bi_ISAC}. Suppose we want to track user $k$ located at $[x~~y]^T$.\footnote{Tracking for multiple users can be done sequentially as in \cite{Beam_space_processing}.} During each measurement epoch $l$ spaced $\Delta T$ apart, each receiver $i$ first estimates the user $k$ location as described in Section \ref{subsection:Para_est}. Denote this estimate as $\boldsymbol{T}_i^l = [\tilde{x}_i^l ~~ \tilde{y}_i^l]^T$. $N_{\text{sel}}$ receivers with the lowest GDOP are selected and their estimates are fused, as mentioned in \ref{sec:fusion} to get the position estimate ${\boldsymbol{T}^l} = [\tilde{x}^l ~~ \tilde{y}^l]^T$. Finally, the prediction of the next location is based on the kinematic model in \cite{Beam_space_processing} that uses the last 3 locations. The next predicted location ${\boldsymbol{T}}_{\text{pred}}^{l+1} = [\hat{x}^{l+1}~~\hat{y}^{l+1}]^T$ and predicted AoD $\hat{\theta}^{l+1}$ are obtained assuming constant acceleration, during the last 3 measurements  as follows: 
\begin{align*}
\hat{x}^{l+1} \approx 3 \tilde{x}^l - 3 \tilde{x}^{l-1} + \tilde{x}^{l-2}, \\  
\hat{y}^{l+1} \approx 3 \tilde{y}^l - 3 \tilde{y}^{l-1} + \tilde{y}^{l-2}, \\  
\hat{\theta}^{l+1} = \arctan{(\hat{y}^{l+1}/\hat{x}^{l+1})}.
\end{align*}
Any error in the estimated location propagates to subsequent measurement epochs through the kinematic model. To limit propagation of large errors, we used circular gating at each receiver to limit the error propagation, i.e.,  when the estimated location $\boldsymbol{T}_i^l$ is outside the gating circle of radius $\beta$ (chosen based on $v_{max}$) from the last predicted location $\boldsymbol{T}_{\text{pred}}^l$, we declare it as an invalid estimate or miss. If all the receivers become invalid, we use the current predicted location for the subsequent kinematic prediction model. When there are 3 consecutive misses for the same user, its tracking ceases. 

\begin{table}[t]
    \caption{Parameters used in the simulation \cite{Beam_space_processing}}
    \label{tab:sim_para}
    \begin{minipage}{\columnwidth}
    \centering
    \begin{tabular}{| c | c |}
        \hline
        $N_{t}$ = $N_{r}$ = 64 & 
        $f_{c}$ = 60~GHz \\
          \hline
        $\Delta f$ = 1~MHz&
        $d_{max}$= 100~m \\
          \hline
        $v_{max}$ = 30~m/s &
        $\sigma_{rcs}$ = 20 ~dBsm \\
          \hline
        $N_{o}$ = $2 \times 10^{-21}$  ~W/Hz & $\Delta T$ = 100 ms \\
            \hline
        No. of runs = 100 & $\beta = 6$~m\\
        \hline
        $M = 512$, $N = 64$ & $S = 1$\\
        \hline
    \end{tabular}
    \end{minipage}
\label{tab:Bi-ISAC-para}
\end{table}


\begin{figure}[t]
\centering
\input{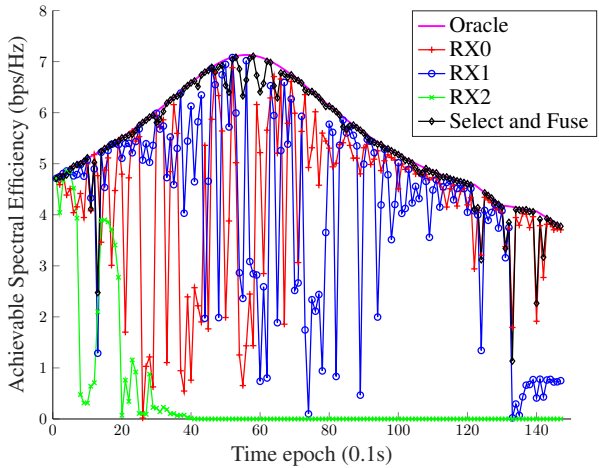}
\caption{Achievable Spectral efficiency vs Time step for the User path for $M = 512$, $P_{T} = 5$ dBm}
\label{fig:SP_M_512}
\end{figure}

\section{Simulation Results}
\label{sec:simulations}

In this section, we present simulation results to show the performance of the proposed beam tracking scheme. We consider the road trajectory of a user vehicle (Target) as shown in Fig.~\ref{fig:path2}.This trajectory is chosen only for illustrating the main results. The proposed techniques work for other trajectories as well. The location of the base station (TX) and the three sensing receivers RX0, RX1, and RX2 are also shown. The echoes of the communication/ISAC signal from the TX to the target are processed by receivers (RX0, RX1, and RX2) to estimate the user's location. A circular gating condition is evaluated to find the receivers with valid estimates. Among the receivers with valid estimates, two receivers with the lowest GDOP are selected. The selected receiver's estimates are fused centrally in the fusion centre. The next location of the target is predicted using the kinematic model in \cite{Beam_space_processing} and the beamforming weights for the next communication phase are sent to the TX enabling accurate beamforming. The downlink communication channel parameters used in the simulation are given in Table~\ref{tab:sim_para}. The beamforming performance is measured by calculating the achievable  Spectral Efficiency (SE) for the user $k$ with a single antenna as in \cite[eqn. (16)]{Beam_space_processing}
\begin{equation}
    SE = \mathbb{E} \left[ \log_{2} \left(1+ \left(\frac{\lambda}{4 \pi d_{k1} }\right)^2 
           \frac{P_{T} |\boldsymbol{a}^{H}(\theta_{k}) \boldsymbol{f}(\Hat{\theta}_{k})  |^{2}}{K N_{0} M (\Delta f)} \right)   \right]. \nonumber
\end{equation}
By reducing the Predicted AoD Error i.e.~$\theta - \Hat{\theta}$ (PAE), the TX can focus the communication beam accurately toward the user, thereby increasing the SE.

In Fig.~\ref{fig:SP_M_512}, the achievable Spectral Efficiency (SE) as the user moves along the path in Fig. \ref{fig:path2} is plotted for an ISAC waveform with 512 subcarriers ($M = 512$). The following 5 schemes are compared: (1) Location estimated only by RX0 for beam tracking, (2) Location estimated only by RX1 for beam tracking, (3) Location estimated only by RX2 for beam tracking, (4) Location estimated by selection and fusion of location estimates of 2 out of 3 receivers, and (5) Exact user's location provided by an oracle for beam tracking. We also implemented the monostatic system as in \cite{Beam_space_processing} and its performance is close to the oracle performance. However, the monostatic system requires full isolation between the TX and RX antennas which requires advanced full-duplex architectures for self-interference cancellation.
It should be noted that the estimation error in AoD for beamforming depends heavily on the geometry of the TX-target-RX location. It can be seen in Fig. \ref{fig:SP_M_512} that the spectral efficiency can be lower even when the receiver is closer to the target because of the geometry. When the target is in the turning region (highlighted by the black box in Fig.~\ref{fig:path2}) corresponding to the duration between the $40^{th}$ and $75^{th}$ time epochs in Fig.~\ref{fig:SP_M_512}, there is a pronounced drop in SE for the cases where a single receiver is used for location tracking (RX0 or RX1).
For the case of RX2, once the targets entered the turn region, the estimation error increases more rapidly and tracking of the target fails (Hence the SE goes down to zero).
  This is because, as explained in \cite{Why_mono_better}, in a bistatic configuration, when the target approaches the baseline (defined as the line connecting the transmitter (TX) and receiver (RX2)), the error in positional estimate increases, i.e., GDOP$_k = \sqrt{\text{trace}(\boldsymbol{\Sigma}_k)}$ increases.  
In the same region, the performance of the fusion system based on (\ref{equ:fuse_meas}) (Fusion) is significantly better and almost identical to the performance with oracle location information. A lower average GDOP over time results in higher average spectral eficiency.
\begin{figure}[t]
\centering
\input{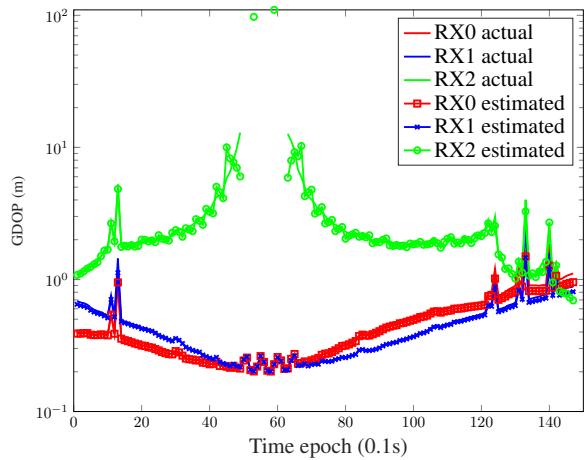}
\caption{GDOP vs Time step for the User path for $M = 512$, $P_{T} = 5$ dBm}
\label{fig:GDOP}
\end{figure}

\begin{figure}[t]
\centering
\input{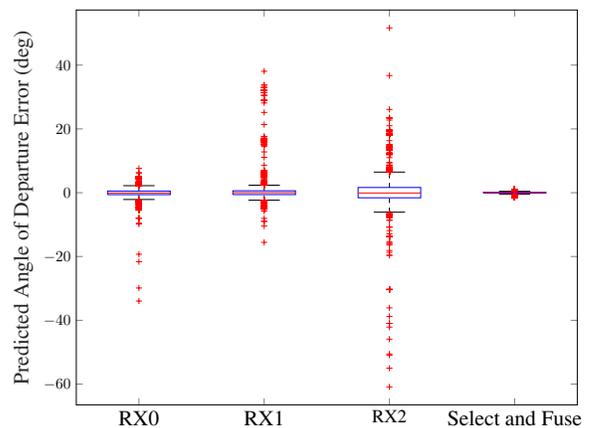}
\caption{Box plot of Predicted AoD Error (deg)}
\label{fig:Ang_err_box}
\end{figure}

For the system which select and fuses the individual estimates from the receivers, their individual GDOP is plotted in Fig.~\ref{fig:GDOP}.
The GDOP calculation is based on two methods: (1) Using  $\rho_{0}$ in equation (\ref{rho_0_actual}) and $\rho_{1} = \left| \boldsymbol{w}^{H}(\Hat{\phi})\boldsymbol{b}(\phi)\right|^{2} \rho_{0}$, with the actual values of $\phi$ and $\theta$ (referred as 'actual'), (2) Using estimated values - $\rho^{est}_{0}$ and $\rho^{est}_{1}$ using equations ($\ref{rho_0_est}), (\ref{roh_1_est}$). We observe from Fig. \ref{fig:GDOP}, that the estimated GDOP closely follows the actual GDOP. The sudden spikes in the GDOP plots are caused by the discontinuity of the user's path. For receivers RX0 and RX1, the GDOP is small compared to RX2. The small wiggle near the $60^{th}$ epoch is due to the inadequacy of our prediction model to track the semicircular path. For the RX2 receivers when the target approaches the baseline, the GDOP increases rapidly as explained in \cite{Why_mono_better}. Overall, it is clear that geometry plays a vital role in the case of bistatic setup and it is always better to fuse the estimates for better tracking, whenever possible.
\begin{figure}[t]
\centering
%
%
\definecolor{mycolor1}{rgb}{1.00000,0.00000,1.00000}%
\begin{tikzpicture}  [scale=0.57]

\begin{axis}[%
width=4.521in,
height=3.566in,
at={(0.758in,0.481in)},
scale only axis,
xmin=-5,
xmax=20,
xlabel style={font=\color{white!15!black}},
xlabel={\Large{Transmitted Power $P_T$ (dBm)}},
ymin=0,
ymax=12,
ylabel style={font=\color{white!15!black}},
ylabel={\Large{Average Achievable Spectral Efficiency (bps/Hz)}},
axis background/.style={fill=white},
legend style={at={(0.10,0.673)}, anchor=south west, legend cell align=left, align=left, draw=white!15!black}
]
\addplot [color=mycolor1, very thick]
  table[row sep=crcr]{%
-5	2.53742786151778\\
-4	2.80848690199122\\
-3	3.08959755886174\\
-2	3.37943635849265\\
-1	3.67675318311993\\
0	3.98040058533325\\
1	4.28935090478467\\
2	4.60270283477074\\
3	4.91967978032415\\
4	5.239622497447\\
5	5.56197826688566\\
6	5.88628841765033\\
7	6.21217552380996\\
8	6.53933114766974\\
9	6.86750463832958\\
10	7.19649322684881\\
11	7.52613347815118\\
12	7.85629404769869\\
13	8.18686962918804\\
14	8.51777595191883\\
15	8.84894568051376\\
16	9.18032507633447\\
17	9.51187129314738\\
18	9.84355019546905\\
19	10.1753346042284\\
20	10.5072028896331\\
};
\addlegendentry{\Large{Oracle}}

\addplot [color=red, very thick, mark size=2.0pt, mark=+, mark options={solid, red}]
  table[row sep=crcr]{%
-5	0.536751103833994\\
-4	0.866765729443609\\
-3	1.17621369487377\\
-2	1.22901011556439\\
-1	1.30710629563572\\
0	1.90583411350748\\
1	2.62777464307793\\
2	3.2791241395311\\
3	3.74400964684755\\
4	3.94200129655146\\
5	4.35766539562266\\
6	4.54955351248764\\
7	4.92106010504849\\
8	5.23866959434012\\
9	5.55535738481627\\
10	5.88758849304046\\
11	6.28384590850307\\
12	6.62906613351875\\
13	6.96066700555614\\
14	7.28840953345455\\
15	7.61218098783383\\
16	7.93837732395616\\
17	8.26247403278717\\
18	8.58932663270863\\
19	8.91675217457413\\
20	9.2458479188596\\
};
\addlegendentry{\large{RX0}}

\addplot [color=blue, very thick, mark size=2.0pt, mark=o, mark options={solid, blue}]
  table[row sep=crcr]{%
-5	0.605419801202588\\
-4	0.710669512562055\\
-3	1.20858732476414\\
-2	1.55985443843687\\
-1	2.03135520969798\\
0	2.49989707457558\\
1	2.87272459943988\\
2	3.23905899840942\\
3	3.51896970598514\\
4	3.8075380378974\\
5	4.01132815097419\\
6	4.38924614937027\\
7	4.66806786965381\\
8	4.92262020397579\\
9	5.22428362861845\\
10	5.57074831933406\\
11	5.86690163623929\\
12	6.35507178480805\\
13	6.71983643286802\\
14	7.08633322736142\\
15	7.41824163812577\\
16	7.77619306790329\\
17	8.05717996777355\\
18	8.33556901438624\\
19	8.73576853533252\\
20	9.09873461703476\\
};
\addlegendentry{\Large{RX1}}

\addplot [color=green, very thick, mark size=2.0pt, mark=x, mark options={solid, green}]
  table[row sep=crcr]{%
-5	0.0862811122260589\\
-4	0.103749034807002\\
-3	0.127719657547606\\
-2	0.164439949811817\\
-1	0.201866699642465\\
0	0.212860412765746\\
1	0.2457528450611\\
2	0.311364478968143\\
3	0.362232035407093\\
4	0.369125154343885\\
5	0.414841447942671\\
6	0.444398300431865\\
7	0.559684979825016\\
8	0.612628651871764\\
9	0.701734903022653\\
10	0.7773415490391\\
11	0.844397204837834\\
12	0.916001794978557\\
13	0.990650653949656\\
14	1.09500998635069\\
15	1.15314199539326\\
16	1.21512861883623\\
17	1.35145681675709\\
18	1.43863523244532\\
19	1.54165239460484\\
20	1.65860614602011\\
};
\addlegendentry{\Large{RX2}}

\addplot [color=black, very thick, mark size=3.5pt, mark=diamond, mark options={solid, black}]
  table[row sep=crcr]{%
-5	2.3479692465233\\
-4	2.56510631213219\\
-3	2.88085966677805\\
-2	3.18015815163917\\
-1	3.46396941422482\\
0	3.76175229698041\\
1	4.06345921081381\\
2	4.36727834273774\\
3	4.67310506057933\\
4	4.98662264167326\\
5	5.29803865785385\\
6	5.61619674850948\\
7	5.93213088111233\\
8	6.25246082169812\\
9	6.57357961916265\\
10	6.89576955809929\\
11	7.21705919566424\\
12	7.54028118595329\\
13	7.86429334443383\\
14	8.18789455806458\\
15	8.51175624878353\\
16	8.83637638646842\\
17	9.16115166316772\\
18	9.48621881063338\\
19	9.81122798089132\\
20	10.1362603823559\\
};
\addlegendentry{\Large{Select and Fuse}}

\end{axis}

\begin{axis}[%
width=5.833in,
height=4.375in,
at={(0in,0in)},
scale only axis,
xmin=0,
xmax=1,
ymin=0,
ymax=1,
axis line style={draw=none},
ticks=none,
axis x line*=bottom,
axis y line*=left
]
\end{axis}
\end{tikzpicture}%
\caption{Average Achievable Spectral Efficiency for the User path for $M = 512$ using Fully Digital Receivers}
\label{fig:Avg_SP_M_512}
\end{figure}
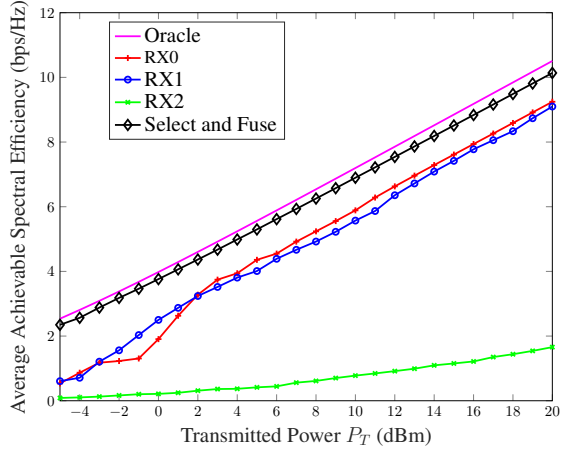

\begin{figure}[t]
\centering
%
%
\definecolor{mycolor1}{rgb}{1.00000,0.00000,1.00000}%
\begin{tikzpicture}[scale = 0.57]

\begin{axis}[%
width=4.521in,
height=3.566in,
at={(0.758in,0.481in)},
scale only axis,
xmin=-5,
xmax=20,
xlabel style={font=\color{white!15!black}},
xlabel={\Large{Transmitted Power $P_T$ (dBm)}},
ymin=0,
ymax=12,
ylabel style={font=\color{white!15!black}},
ylabel={\Large{Average Achievable Spectral Efficiency (bps/Hz)}},
axis background/.style={fill=white},
legend style={at={(0.1,0.68)}, anchor=south west, legend cell align=left, align=left, draw=white!15!black}
]
\addplot [color=mycolor1, very thick]
  table[row sep=crcr]{%
-5	2.53742786151778\\
-4	2.80848690199122\\
-3	3.08959755886174\\
-2	3.37943635849265\\
-1	3.67675318311993\\
0	3.98040058533325\\
1	4.28935090478467\\
2	4.60270283477074\\
3	4.91967978032415\\
4	5.239622497447\\
5	5.56197826688566\\
6	5.88628841765033\\
7	6.21217552380996\\
8	6.53933114766974\\
9	6.86750463832958\\
10	7.19649322684881\\
11	7.52613347815118\\
12	7.85629404769869\\
13	8.18686962918804\\
14	8.51777595191883\\
15	8.84894568051376\\
16	9.18032507633447\\
17	9.51187129314738\\
18	9.84355019546905\\
19	10.1753346042284\\
20	10.5072028896331\\
};
\addlegendentry{\Large{Oracle}}

\addplot [color=red, very thick, mark size=2.0pt, mark=+, mark options={solid, red}]
  table[row sep=crcr]{%
-5	0.982408766330181\\
-4	1.32212667866684\\
-3	1.52968973806436\\
-2	1.89083048950371\\
-1	2.0793247224973\\
0	2.47691692891124\\
1	2.77088355978236\\
2	3.2298482195564\\
3	3.50358414424106\\
4	3.79894120836159\\
5	4.07999445554329\\
6	4.57048109116126\\
7	4.78378516982562\\
8	5.17427851316601\\
9	5.41291659141286\\
10	5.87014587234676\\
11	6.19006536471094\\
12	6.52121889353054\\
13	6.75538173760758\\
14	7.23633229265231\\
15	7.46710492473348\\
16	7.8821650279942\\
17	8.2597958085128\\
18	8.46469014510647\\
19	8.96645476901664\\
20	9.30292413924337\\
};
\addlegendentry{\Large{RX0}}

\addplot [color=blue, very thick, mark size=2.0pt, mark=o, mark options={solid, blue}]
  table[row sep=crcr]{%
-5	1.41762558584091\\
-4	1.68194359929055\\
-3	1.86913127697518\\
-2	2.1105665325276\\
-1	2.3654337152847\\
0	2.80535463845475\\
1	3.04821068835198\\
2	3.3408891119222\\
3	3.57336932455855\\
4	3.93547957691682\\
5	4.27829725659235\\
6	4.57823978534658\\
7	4.93846787576095\\
8	5.24334021100743\\
9	5.51841336945911\\
10	5.80798620179985\\
11	6.19467075290737\\
12	6.54316347568618\\
13	6.79258546768379\\
14	7.08749221552064\\
15	7.47761217510664\\
16	7.68033797292239\\
17	8.15545623601561\\
18	8.47459617119056\\
19	8.84276619112624\\
20	9.16703165766879\\
};
\addlegendentry{\Large{RX1}}

\addplot [color=green, very thick, mark size=2.0pt, mark=x, mark options={solid, green}]
  table[row sep=crcr]{%
-5	0.145833695039606\\
-4	0.175769127121052\\
-3	0.185657339730306\\
-2	0.228379844694226\\
-1	0.250235995410161\\
0	0.29454172964804\\
1	0.331018725014932\\
2	0.348232104691457\\
3	0.391408728929425\\
4	0.433473519442702\\
5	0.489191347220852\\
6	0.528553751976103\\
7	0.589925872689164\\
8	0.643658768097992\\
9	0.708651748271637\\
10	0.772507744300063\\
11	0.817662157071882\\
12	0.896463258055545\\
13	0.937970701966371\\
14	1.04985958265044\\
15	1.16001833562545\\
16	1.24148579881075\\
17	1.35207395077555\\
18	1.45122245359298\\
19	1.51026235725457\\
20	1.61506585957141\\
};
\addlegendentry{\large{RX2}}

\addplot [color=black, very thick, mark size=3.5pt, mark=diamond, mark options={solid, black}]
  table[row sep=crcr]{%
-5	2.42664631132502\\
-4	2.69492440407513\\
-3	2.96593737042877\\
-2	3.24667302115147\\
-1	3.53549270623805\\
0	3.82916994168247\\
1	4.12885423358925\\
2	4.43362935237849\\
3	4.74166531346037\\
4	5.05333856989014\\
5	5.36726337470789\\
6	5.68355326545226\\
7	6.00170432917596\\
8	6.3211267962199\\
9	6.64187096633108\\
10	6.96350307376245\\
11	7.28600466362205\\
12	7.6092491791157\\
13	7.93296991482228\\
14	8.25659521635385\\
15	8.58092349845253\\
16	8.90538790046255\\
17	9.23016726942699\\
18	9.55504074041151\\
19	9.88008063447569\\
20	10.2051412749847\\
};
\addlegendentry{\Large{Select and Fuse}}

\end{axis}

\begin{axis}[%
width=5.833in,
height=4.375in,
at={(0in,0in)},
scale only axis,
xmin=0,
xmax=1,
ymin=0,
ymax=1,
axis line style={draw=none},
ticks=none,
axis x line*=bottom,
axis y line*=left
]
\end{axis}
\end{tikzpicture}%
\caption{Average Achievable Spectral Efficiency for the User path for $M = 512$ using HDA Receivers}
\label{fig:Avg_SP_M_512_HDA}
\end{figure}
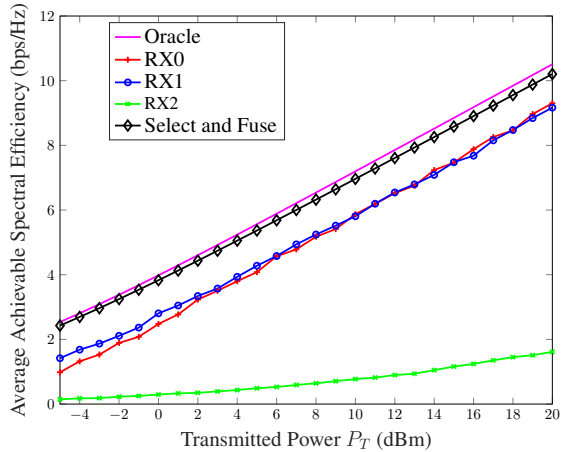

In Fig.~\ref{fig:Ang_err_box}, we plot the Predicted AoD Error (PAE)  in degrees for various system configurations.
The Fusion of estimates results in lesser PAE when compared to a single receiver configuration. The higher the width of the box, the higher the PAE.
In Fig.~\ref{fig:Avg_SP_M_512}, We plot the time average achievable spectral efficiency over the whole path for various transmitted powers. It can be observed that the overall SE is significantly better for the Fusion receiver compared to the individual receivers. 

Finally, we consider a Hybrid Digital Analog (HDA) architecture with only $N_r^{rf} = 4$ RF chains for the 64 antenna receiver. This significantly reduces complexity compared to the fully digital architecture. We modify the HDA architecture receiver in \cite{Beam_space_processing} and adapt it to our bistatic configuration and used Time-Half bandwidth product (THBW =$1$) for generation of the reduction matrix. The performance results in Fig.~\ref{fig:Avg_SP_M_512_HDA} show that the fusion of estimates from HDA receivers is significantly better than the respective individual single HDA receivers. Even with HDA receivers, the performance with fusion is close to the performance with Oracle location information.
\section{Conclusion} 
\label{sec:conclusions}
We proposed a simultaneous communication and beam tracking scheme based on receiver selection and maximum likelihood fusion of the selected multiple bistatic measurements based on reflected ISAC/communication signals. Appropriate error covariance matrices required for selection and fusion were estimated from the received signal. In bistatic sensing, the TX-target-RX geometry plays an important role in estimation accuracy and this is captured by our estimated error covariance matrices. Using the proposed scheme, the average spectral efficiency approaches that of a system with perfect user location knowledge. The proposed technique can also be applied to systems with hybrid digital-analog architectures. 
Compared to a monostatic simultaneous communication and tracking system \cite{Beam_space_processing} which requires a complex design to suppress self-interference, our method works with simple half-duplex digital receivers. Our future work will focus on extending our approach to multi-user scenarios where users are not well separated.

\bibliographystyle{IEEEtran}
\bibliography{ref}

\end{document}